\documentclass[useAMS,usenatbib]{mn2e}
\usepackage{epsfig}
\usepackage{epstopdf}
\usepackage{graphicx}
\usepackage{amssymb}
\usepackage{amsmath}
\usepackage{graphicx}
\usepackage{subcaption}
\usepackage[utf8]{inputenc}
\usepackage{ulem}
\usepackage[export]{adjustbox}
\usepackage{wrapfig}
\usepackage{textcomp}
\usepackage[usenames,dvipsnames]{color}
\usepackage{gensymb}
\usepackage{booktabs}
\usepackage{amsmath}
\usepackage{lscape}
\usepackage[english]{babel}
\usepackage{graphicx}
\usepackage{float}
\usepackage{fixltx2e}

\usepackage{hyperref} 
\hypersetup{colorlinks=true, linkcolor=blue, citecolor=blue, filecolor=blue, urlcolor=black}

\begin{document}

\title[Clustering in Horizon-AGN]{Comparing Galaxy Clustering in  Horizon-AGN Simulated Lightcone Mocks and VIDEO Observations}
\author[Peter Hatfield]{P. W. Hatfield$^{1}$\thanks{peter.hatfield@physics.ox.ac.uk}, C. Laigle$^{2}$, M.J. Jarvis$^{2, 3}$, J.Devriendt$^{2}$, I.Davidzon$^{4}$,\newauthor O.Ilbert$^{5}$, C. Pichon$^{6,7,8}$, Y. Dubois$^{6}$\\
$^{1}$Department of Physics, Clarendon Laboratory, University of Oxford, Parks Road, Oxford OX1 3PU, UK\\
$^{2}$Astrophysics, University of Oxford, Denys Wilkinson Building, Keble Road, Oxford, OX1 3RH, UK\\
$^{3}$Department of Physics, University of the Western Cape, Bellville 7535, South Africa\\
$^{4}$IPAC, Mail Code 314-6, California Institute of Technology, 1200 East California Boulevard, Pasadena, CA 91125, USA\\
$^{5}$Aix Marseille Univ, CNRS, LAM, Laboratoire d'Astrophysique de Marseille, Marseille, France\\
$^{6}$Sorbonne Universit\'es, CNRS, UMR 7095, Institut d'Astrophysique de Paris, 98 bis bd Arago, 75014 Paris\\
$^{7}$Institute for Astronomy, University of Edinburgh, Royal Observatory, Blackford Hill, Edinburgh, EH9 3HJ, United Kingdom\\
$^{8}$Korea Institute for Advanced Study (KIAS), 85 Hoegiro, Dongdaemun-gu, Seoul, 02455, Republic of Korea\\
}


\date{2019}

\pagerange{\pageref{firstpage}--\pageref{lastpage}} \pubyear{2019}

\maketitle

\label{firstpage}

\begin{abstract}

Hydrodynamical cosmological simulations have recently made great advances in reproducing galaxy mass assembly over cosmic time - as often quantified from the comparison of their predicted stellar mass functions to observed stellar mass functions from data. In this paper we compare the \textit{clustering} of galaxies from the hydrodynamical cosmological simulated lightcone {\sc Horizon-AGN}, to clustering measurements from the VIDEO survey observations. 
Using mocks built from a VIDEO-like photometry, we first explore the bias introduced into clustering measurements by using stellar masses and redshifts derived from SED-fitting, rather than the intrinsic values. The propagation of redshift and mass statistical and systematic uncertainties in the clustering measurements causes us to underestimate the clustering amplitude. We find then that clustering and halo occupation distribution (HOD) modelling results are qualitatively similar in {\sc Horizon-AGN} and VIDEO. However at low stellar masses {\sc Horizon-AGN} underestimates the observed clustering by up to a factor of $\sim$3, reflecting the known excess stellar mass to halo mass ratio for {\sc Horizon-AGN} low mass haloes, already discussed in previous works. This reinforces the need for stronger regulation of star formation in low mass haloes in the simulation. Finally, the comparison of the stellar mass to halo mass ratio in the simulated catalogue, inferred from angular clustering, to that directly measured from the simulation, validates HOD modelling of clustering as a probe of the galaxy-halo connection.

\end{abstract}


\section{Introduction}

Parallel to the success of wide-field surveys in \textit{observing} large numbers of galaxies, there have also been great advances in \textit{simulating} large numbers of galaxies in cosmological simulations (see \citealp{Somerville2014} for a review). Cosmological simulations typically model a fraction of the Universe as a finite cube with periodic boundary conditions, take some set of initial conditions, some formulation of the physics of the Universe that the simulator is interested in capturing, and let this realization of the universe evolve from the beginning of time to $z=0$. Critically, simulations are the only way to solve the full growth of cosmological structure into the strongly non-linear regime, including baryonic physics (down to the resolution limit of the simulation) - the analytic derivation being possible only for dark matter (DM) in the linear or weakly non-linear regime (see e.g. \citealp{Peebles1980}). Cosmological simulations have dramatically changed over the last $\sim 50$ years due to the improvement of computing power, from being only able to qualitatively capture the large-scale structure of the Universe, to being able to capture in great detail a wide range of baryonic physics processes.

Galaxy populations can be either 'grafted' on a DM $N$-body simulation according to some prescriptions  (e.g. {\sc Galform} \citealp{Bower2005}; {\sc SAGE} \citealp{Croton2006}, \citealp{Croton2016}; {\sc SeSAM} \citealp{Neistein2009}; the model of \citet{Guo2011} in the Millennium simulation \citealp{Springel2005}), or simulated directly by following the physics of gas and star-formation in hydrodynamical simulations (e.g. {\sc Horizon-AGN} \citealp{Dubois2014}; EAGLE, \citealp{Schaye2015,McAlpine2015}; Illustris \citealp{Vogelsberger2014}; Mufasa \citealp{Dave2016}; MassiveBlack-II \citealp{Khandai2014}). The first method, called semi-analytic models (SAM), provides the opportunity to populate very large  volumes at low computational expense, but relies on a large number of tuneable parameters, as the rich diversity of galaxy properties has to be inferred from the DM halo mass and merger trees only. Hydrodynamic simulations, in contrast, seek to incorporate physics in a more fundamental way down to the smallest possible scale by consistently following simultaneously DM, gas and stars - but are therefore vastly more computationally expensive, and are limited to smaller volumes (typically several hundred of Mpc for the largest ones).

No simulation can capture all physics present in the real Universe. Instead we wish to investigate how successful a simulation is in recreating observations and understand which aspects of galaxy formation physics are correctly reproduced, and which aspects are currently missing. One approach is to use 1-point statistics as a measure of how successful a simulation is at recreating the observed stellar mass function  (see e.g. \citealp{Schaye2015}, \citealp{Kaviraj2017}), star formation rates (e.g. \citealp{Katsianis2015}), morphology (e.g. \citealp{Dubois2016}) and scaling relations (e.g. \citealp{Jakobs2018}).  1-point statistics are however not sufficient to constrain all aspects of galaxy mass assembly. The spatial distribution of the galaxies, as quantified from 2-point statistics, gives extra information e.g. on the scale at which baryonic processes are important (e.g. \citealp{Chisari2018}), how galaxies interact which each other and their environment (see e.g. \citealp{Li2011} for the comparison of spectroscopic observations and SAMs) and the galaxy-halo connection (\citealp{Wechsler2018}). Comparing the clustering of galaxies in observations to galaxy clustering in hydrodynamic simulations therefore provides additional constraints on galaxy evolution models (see e.g. \citealp{Saghiha2016}) - this is the focus of  this work.

Most previous comparisons between clustering in observations and in simulations have focused on galaxy bias (e.g. \citealp{Blanton1998,Blanton1999,Cen1998}, \citealp{Yoshikawa2001}, \citealp{Weinberg2004c}), the estimation of which is pivotal for making cosmological inferences from these measurements. From the point of view of galaxy evolution however, the bias does not include all information about the spatial distribution of galaxies especially at non-linear scales. In order to constrain the galaxy evolution model, especially baryonic feedback, we need to probe these small-scales, which matter the most for galaxy formation and evolution. These small scales are often modelled with a Halo Occupation Distribution (HOD) phenomenology. Small-scale clustering observations (and corresponding HOD inferences) have been compared to simulation predictions of clustering to test galaxy evolution models in the literature e.g. \citet{McCracken2007} compare clustering in observations and mock catalogues from semi-analytic models down to scales of $\sim0.1$Mpc in the COSMOS field.  \citet{Farrow2015} more recently compared clustering in a $\sim 180$deg$^2$ survey to clustering predictions from {\sc Galform} down to sub-Mpc scales at $z<0.5$. They found that  {\sc Galform} correctly predicts the well-known trend of more massive galaxies being more strongly clustered. However, the model made incorrect predictions for other aspects of clustering, in particular the correlation function on the smallest scales, and the clustering of the most luminous galaxies. The impact of observational biases on clustering measurements, and the validity of HOD modelling (how accurate the inferences made from the approach are) have also been examined in the literature. \citet{Crocce2016a} compare how inferences from clustering differ when different photometric redshift calculation approaches are used (machine learning based versus template based). \citet{Beltz-Mohrmann2019a} test the validity of HOD inferences by modelling the clustering in hydrodynamical simulations, and then applying the resulting occupation model to a dark matter only simulation. They found that the different halo mass functions (due to the presence of baryons) between the hydrodynamical simulations and the dark matter only simulation caused some differences in the resulting clustering - but that small corrections could largely account for this effect (at least for their high luminosity sample). Finally, \citet{Zentner2014} (see also \citealp{Hearin2015}) present concerns that `assembly bias' (the phenomenon that the clustering of dark matter haloes depends on assembly time, as well as mass) could seriously bias inferences from HOD modelling (which typically only includes dependences on halo mass). They suggest that halo mass estimates could be biased by up to $\sim0.5$ dex if assembly bias is not incorporated, more than conventional uncertainty calculations would suggest - a serious concern.

In this work we seek to build on previous work comparing observations and hydrodynamical simulations (see e.g. \citealp{Artale2016} in EAGLE at $z\sim0.1$, \citealp{Springel2017} in Illustris), by comparing clustering and HOD measurements (as a function of stellar mass) between observations at $z>0.5$ and hydrodynamic cosmological simulations. On the simulation side, we work with the {\sc Horizon-AGN} simulation, a hydrodynamic simulation with gas cooling, star formation and stellar and AGN feedback prescriptions (\citealp{Dubois2014}). The focus is on the comparison of  clustering measurements and HOD modelling between VIDEO observations and mock catalogues built from the {\sc Horizon-AGN} lightcone (\citealp{Laigle2019}).

Our goal is three-fold. Firstly, by using mock catalogues built from the simulation, both including and not including the observational systematics\footnote{Photometric uncertainties and systematics arising when deriving the masses and redshifts from SED-fitting}, we can quantify how the propagation of these uncertainties biases the clustering measurements. 
Secondly, the comparison of galaxy clustering in the real and mock VIDEO observations will allow us to constrain the galaxy evolution model implemented in the simulation. Finally the measurement of the galaxy-halo connection, derived from the galaxy-clustering in the mocks compared to the connection measured directly in the simulation will help us to confirm that the HOD model is a good way to infer halo mass.

This paper is organised as follows: we describe the observational data we compare the simulation to in Section \ref{sec:observation_description}. In Section \ref{sec:simulation_description} we briefly describe the {\sc Horizon-AGN} simulation, and the mock catalogue constructed from it.  We then measure the correlation function in the simulation mock catalogue for sub-samples corresponding to those considered in VIDEO in \citet{Hatfield2016} in Section \ref{sec:horizon_results}. We fit HOD models to these simulated observations, and compare and contrast the {\sc Horizon-AGN} results to those from the VIDEO observations (Section \ref{sec:horizon_discussion}). Finally we discuss the implications of our results for understanding which  aspects of galaxy formation physics simulations capture well, and which they don't, and for understanding how confident we can be in inferences from HOD modelling.

Unless specified otherwise, we call ``intrinsic" the quantities (redshifts and masses) which are directly measured from the simulation, and ``photometric" the quantities from the simulation wich are derived through SED-fitting on the mock photometry.

\section{Observations} \label{sec:observation_description}

The VIDEO Survey \citep{Jarvis2013} on the VIRCAM camera on the VISTA telescope (\citealp{Dalton2006}) is a deep wide-field survey covering three fields in the southern hemisphere, each carefully chosen for availability of multiband data, totalling 12 deg$^{2}$. VIDEO is sensitive to similar volumes, redshifts and stellar masses as {\sc Horizon-AGN}, permitting a meaningful comparison between the two. The $5 \sigma$ depths of these VIDEO observations (used in \citealp{Hatfield2016}) in the five bands are $Z=25.7$, $Y=24.5$, $J=24.4$, $H=24.1$ and $K_s=23.8$ for a 2'' diameter aperture.

Our catalogue was constructed by combining the VIDEO data set with data from the T0006 release of the Canada-France-Hawaii Telescope Legacy Survey (CFHTLS) D1 tile \citep{Ilbert2006,Gwyn2012}, which provides photometry with $5 \sigma$ depths (2'' apertures) of $u^{*}= 27.4$, $g^{\prime}=27.9$, $r^{\prime}=27.6$, $i^{\prime}=27.4$ $z^{\prime}=26.1$ over 1 deg$^{2}$ of the VIDEO XMM3 tile. This data set has already been used in many extragalactic studies to date (e.g.  \citealp{White2015,Johnston2015,Hatfield2016}).

The same data processing and sub-sample selection as \citet{Hatfield2016} is used here - see \citet{Hatfield2016} and \citet{Jarvis2013} for a full description of the data set, samples, detection images , detection thresholds and the construction of the SED templates used, but we briefly summarise the reduction process here. The sources were identified using SExtractor, \citep{Bertin1996} source extraction software, with 2'' apertures. Photometric redshifts and stellar masses used in this work were estimated using LePHARE \citep{Arnouts1999,Ilbert2006}, which fits spectral energy distribution (SED) templates to the photometry.

{\sc SExtractor} identified 481,685 sources in the CHFTLS-D1 1~deg$^2$ with detections in at least one band. As described in \citet{Hatfield2016}, we remove sources in regions effected by excess noise and bright stars with a mask. We only include sources with $K_{s}<$23.5, and make a colour cut around a stellar locus, following the approach of \cite{Baldry2010}, to remove stars. VIDEO has a $90$ percent completeness at this depth \citep{Jarvis2013}. \citet{McAlpine2012} estimate that this colour cut leaves stars contributing less than 5 per cent of the sample. The final galaxy sample $\cal{C}_{\rm VIDEO}$ comprises 97,052 sources. The clustering properties of galaxies in this VIDEO-CFHTLS data set have been explored in \citet{Hatfield2016} (where the observed clustering measurements and HOD fits used in this work are taken from) and \citet{Hatfield2017}.

\section{The virtual VIDEO-like catalogue} \label{sec:simulation_description}

\subsection{The {\sc Horizon-AGN} simulation}

The full specifications of {\sc Horizon-AGN} and the details of the physics it incorporates can be found in \citet{Dubois2014}, \citet{Dubois2016}, \citet{Kaviraj2017}, and \citet{Laigle2019}; we give only a brief description here.

{\sc Horizon-AGN} was run with the RAMSES adaptive mesh refinement code \citep{Teyssier2002}. It simulates a box with periodic boundary conditions of comoving width 100 Mpc/h, containing $1024^3$ dark matter particles (mass resolution of $8\times 10^7 \mathrm{M}_{\sun}$),  compatible with a WMAP7 cosmology \citep{Komatsu2010}. Gas dynamics, cooling and heating is followed on the adaptive mesh with a minimum cell size of 1~kpc (constant in physical length). The simulation also follows star formation (star particle mass resolution of $2\times 10^6 \mathrm{M}_{\sun}$), feedback from stars (stellar winds,  type~II and type~Ia supernovae), the evolution of six chemical species, and feedback from AGN\footnote{Although we do not use it in this work, a twin simulation was also run without AGN feedback, the {\sc Horizon-no-AGN} simulation, making it particularly useful for consistently testing the role of AGN in galaxy formation (\citealp{Peirani2016,Beckmann2017}).}.

AGN feedback can be either `radio mode' or `quasar mode' (\citealp{Dubois2012}). Radio mode operates at low black hole accretion rates ($\dot{M}_{\rm BH}/\dot{M}_{\rm Edd}<0.01$, where $\dot{M}_{\rm BH}$ is the black hole accretion rate, and $\dot{M}_{\rm Edd}$ is the the Eddington accretion rate, see \citealp{Castello-Mor2016}), and injects energy into the inter-galactic medium through bipolar jets. Quasar mode operates at higher accretion rates ($\dot{M}_{\rm BH}/\dot{M}_{\rm Edd}>0.01$), and injects energy isotropically.

\subsection{Extraction of galaxies and halos}

A lightcone (a simulated box for which one direction is redshift) 
with opening angle of 1~deg$^{2}$ was constructed on-the-fly from the simulation \citep[see][]{Pichon2010}, with very fine redshift steps (about 22,000 steps up to $z=6$). Such lightcone mimics therefore the geometry of observational surveys. 

Galaxies have been initially identified on this lightcone slices (about 4000 slices up to $z=4$) from the distribution of star particles using the AdaptaHOP structure finder code (\citealp{Aubert2004,Tweed2009}). Structures are selected with a density threshold of 178 times the average matter density and are required to contain at least 50 stellar particles. 

DM haloes have been extracted independently on the same slices from the distribution of DM particles, with a density threshold of 80 times the average matter density and are required to contain at least 100 stellar particles. Galaxies are matched with their closest halos. To this purpose, the centre of the haloes is accurately defined from the shrinking sphere method (\citealp{Power2003}).

\subsection{Production of the photometric catalogues}
A mock photometric catalogue is then generated in order to mimic the VIDEO photometry, following the method presented in \cite{Laigle2019}. We recall below the main aspects of the building of the catalogue.
Galaxy fluxes are calculated from the total distribution of stellar particles in each galaxy using the stellar population synthesis (SPS) models from \citet{Bruzual2003b} with a Chabrier \citep{Chabrier2003} initial mass function (IMF). Each star particle is assumed to behave as a single stellar population, and the total galaxy flux is the sum of the contributions of all star particles. Dust attenuation is implemented along the line-of-sight of each star particle, using  the ${\rm R}_{V}=3.1$ Milky Way dust grain model by \citet{Weingartner2001}.  The gas metallicity distribution around the galaxies is taken as a proxy for dust distribution.

Galaxy spectra are then shifted according to the galaxy redshift, and convolved with the same filter pass-bands as the real data (namely $u^{*}$, $g^{'}$, $r^{'}$, $i^{'}$, $z^{'}$ from CFHTLS, and $Z$, $Y$, $J$, $H$ and $K_{\rm s}$ from VIDEO).  Photometric errors are added in each band to reproduce the $S/N$ distribution and sensitivity limit of VIDEO and CFHTLS filters  (see section \ref{sec:observation_description}). 
From this mock photometry, photometric redshifts and stellar masses are computed using the code {\sc LePhare} \citep{Arnouts2002,Ilbert2006} with a configuration similar to \citealp{Laigle2019}.

Two simulated catalogues are used in this work. One, called $\cal{C}_{\rm Hzagn,sim}$ in the following, contains the intrinsic stellar masses and redshifts. The other, $\cal{C}_{\rm Hzagn,obs}$, uses stellar masses and redshifts derived from the mock photometry through SED-fitting and therefore naturally incorporates the systematics arising when computing physical properties from their photometry. A detailed study of these systematics has been presented in \cite{Laigle2019}. In the following (Section~\ref{sec:observational_biases}), we will in particular investigate how the systematics propagate in the clustering measurements. Note that we incorporate observational limitations only up to a point. In particular, the galaxy photometry is derived from the entire distribution of particles, and not from a two-dimensional extraction on realistic images. Although noise is added to the mock photometry in a statistical way, it does not include systematics which usually degrade the observed photometry, due to blending, object fragmentation, variable PSF, etc. Furthermore, as discussed in \citealp{Laigle2019}, the mock photometry includes less variety than the real one, because of the use of a single and constant IMF, a single SPS model, a single dust attenuation law and the absence of nebular emission lines. As a consequence the uncertainties on the photometric redshifts and stellar masses might be underestimated, as might be the amplitude of the systematics in the clustering measurements  we derive in the following sections.

\begin{figure}
\centering
\includegraphics[scale=0.28]{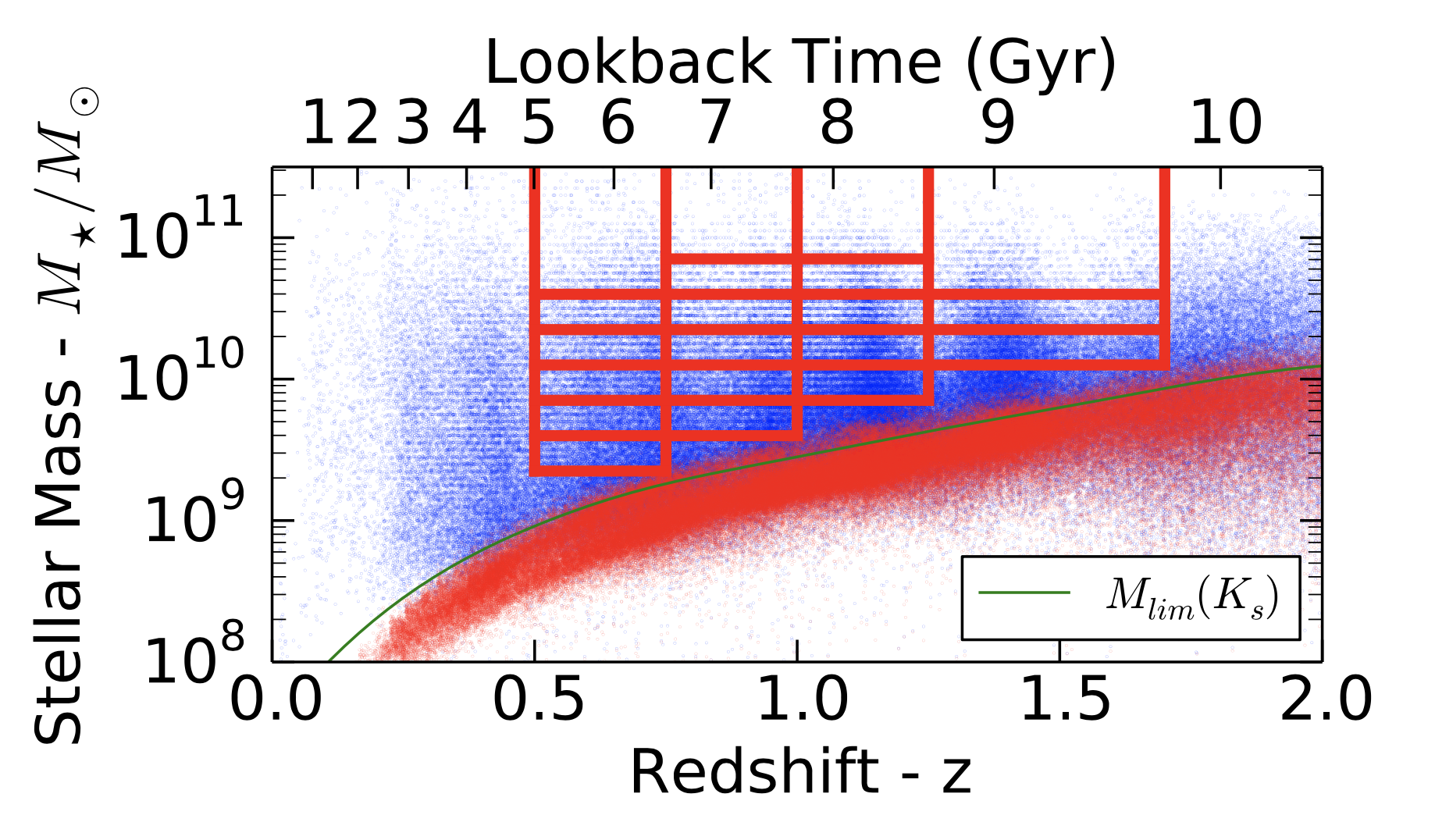}
\caption{Distribution of the stellar masses and redshifts (blue points) of all $K<23.5$ galaxies in $\cal{C}_{\rm Hzagn,obs}$. The red lines denote the bins used in \citet{Hatfield2016}. The red points mark the stellar mass limit for all objects that could be detected with the apparent magnitude limit of $K_{s}<23.5$, and the green curve the implied $90\%$ stellar mass completeness limit, following the approach of \protect\cite{Johnston2015}. The green line matches the bottom edge of the red bins, as a confirmation that the mock catalogue has similar mass limits as $\cal{C}_{\rm VIDEO}$.}
\label{fig:mock_MZ}
\end{figure}

Fig. \ref{fig:mock_MZ} (c.f. Fig.~1 in \citealp{Hatfield2016}) shows the $\cal{C}_{\rm Hzagn,obs}$ masses and redshifts, after having removed galaxies fainter than $K_{\rm s}>23.5$ (based on their mock photometry) as we did with the observations, leaving 259,567 galaxies (more than in VIDEO because of over-production of low-mass galaxies in {\sc Horizon-AGN}, see section \ref{sec:over_production}). We apply the same criteria as \citet{Hatfield2016} from \citet{Pozzetti2010} and \citet{Johnston2015} to determine the 90\% completeness limit. This limit is found by calculating the lowest stellar mass that each galaxy could have been detected at, using $\log_{10}(M_{\mathrm{lim}})=\log_{10}(M_{\star})+0.4(K_{s}-K_{\mathrm{lim}})$ (where $M_{\mathrm{lim}}$ is this lowest mass that the galaxy could have been observed at, $M_{\star}$ is the galaxy mass, $K_{s}$ is the galaxy $K_{s}$-band magnitude, and $K_{\mathrm{lim}}$ is the $K_{s}$-band limit used, $K_{\mathrm{lim}}=23.5$ here). We can then find the 90th percentile of these stellar mass limit values, as a function of redshift, which corresponds to a 90\% completeness limit. This limit is very similar to that found for the observations (as expected), allowing us to consistently compare all the VIDEO subsamples to the {\sc Horizon-AGN} mock catalogue.

\begin{figure}
\centering
\includegraphics[scale=0.45]{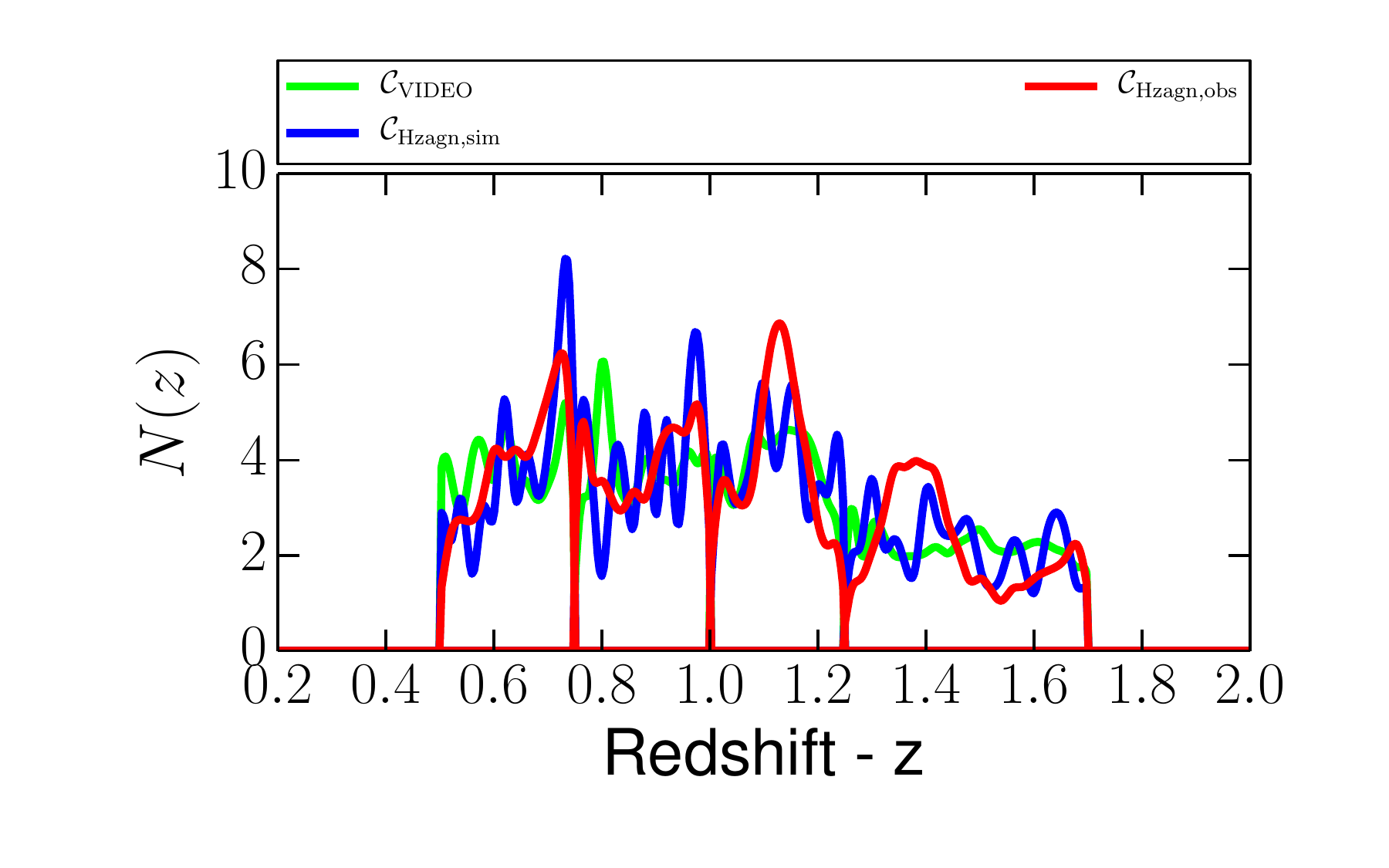}
\caption{The redshift distributions of $\cal{C}_{\rm Hzagn,obs}$, $\cal{C}_{\rm Hzagn,obs}$ and $\cal{C}_{\rm VIDEO}$, each catalogue and each redshift bin normalised to have an area of one.}
\label{fig:redshift_distributions}
\end{figure}

\begin{figure}
\centering
\includegraphics[scale=0.7]{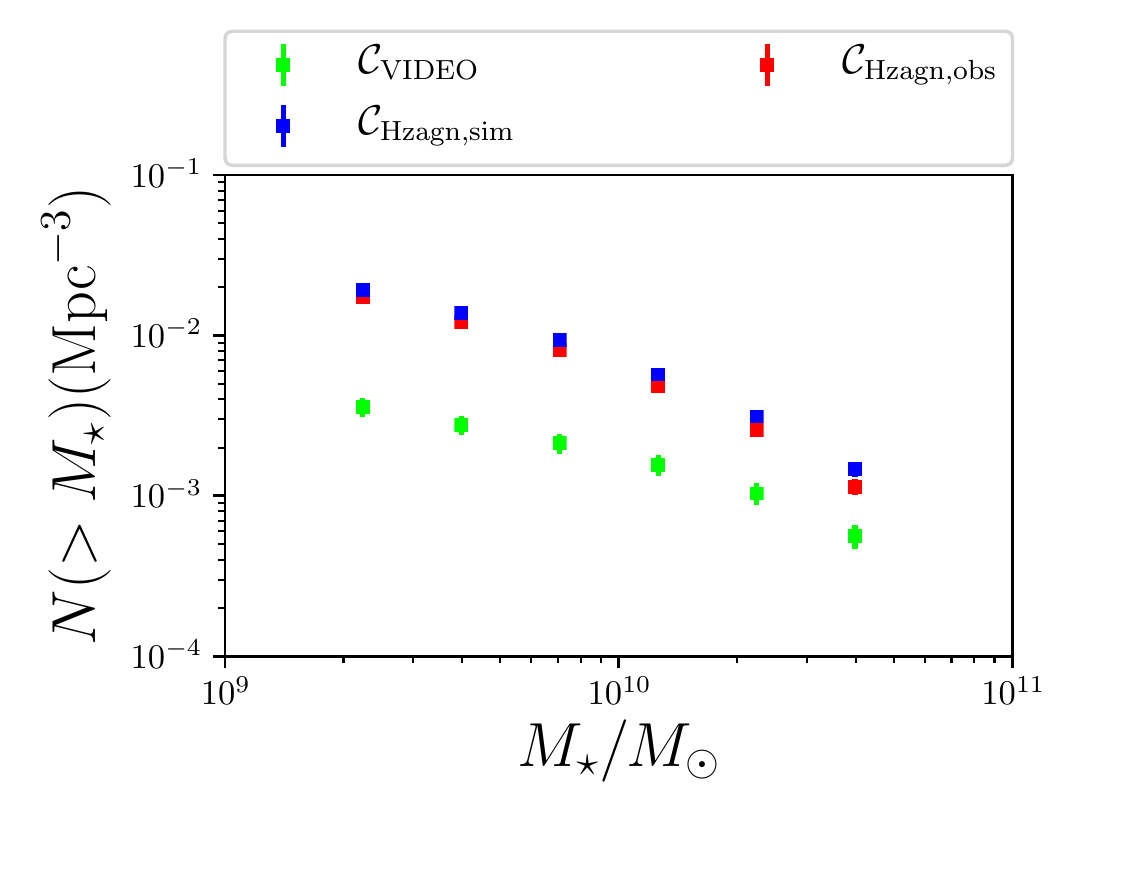}
\caption{The number counts from $\cal{C}_{\rm Hzagn,obs}$, $\cal{C}_{\rm Hzagn,obs}$ and $\cal{C}_{\rm VIDEO}$ for the $0.5<z<0.75$ redshift bin.}
\label{fig:number_counts}
\end{figure}

We also confirm that the redshift distributions within the redshift bins are relatively similar between the two {\sc Horizon-AGN} catalogues and the VIDEO catalogue (figure \ref{fig:redshift_distributions}). The two {\sc Horizon-AGN} catalogues have very simiular redshift distributions (as expected); the medians of the redshift distributions are within 0.002 of each other for the first three redshift bins, and 0.014 different in the fourth redshift bin (the bin with the largest range). Similarly the medians of the redshift distributions in each VIDEO redshift bin are within 0.02 of the {\sc Horizon-AGN} in the first three redshift bins, and 0.04 in the fourth redshift bin. In summary, in terms of comparing like-to-like, in all our analysis we are comparing comparable redshift distributions, with the possible very minor exception of the $1.25<z<1.7$ VIDEO bin, which has a slightly higher median redshift than the corresponding {\sc Horizon-AGN} samples. Finally in figure \ref{fig:number_counts} we also show a sample comparison of the number counts in the three catalogues considered in this work ($\cal{C}_{\rm Hzagn,obs}$, $\cal{C}_{\rm Hzagn,obs}$ and $\cal{C}_{\rm VIDEO}$) as a function of stellar mass. See also figure A3 in \citet{Laigle2019}. The $\cal{C}_{\rm Hzagn,obs}$ and $\cal{C}_{\rm Hzagn,obs}$ 1-point statistics are very similar (note the difference is better interpreted as a shift in the x-axis from a bias in the stellar mass estimates, rather than a y-axis shift in the actual number of galaxies). The $\cal{C}_{\rm VIDEO}$ number counts are substantially different to that in the {\sc Horizon-AGN} catalogues due to different physics in the simulation to in the real Universe - discussed in greater depth in section \ref{sec:over_production}

\section{Methods} \label{sec:methods_description}

In this work we apply the same methodology as used for $\cal{C}_{\rm VIDEO}$  \citep{Hatfield2016} for calculating the angular correlation function and fitting HOD models in both simulated catalogues $\cal{C}_{\rm Hzagn,sim}$ and $\cal{C}_{\rm Hzagn,obs}$. 

Note that unlike in real datasets, there is no need to remove stars and no mask is needed for the simulated data. For $\cal{C}_{\rm Hzagn,obs}$, the redshifts have uncertainties estimated from LePhare and we use the weighting system of \cite{Arnouts2002} when calculating the correlation function; for $\cal{C}_{\rm Hzagn,sim}$ we simply use the true redshifts for each source (equivalently we give each source a weight of one for the redshift range it is in, and zero for the other redshift bins). The VIDEO field and the {\sc Horizon-AGN} lightcone are both 1~deg$^2$ fields, so the integral constraint (a bias to measurements of the correlation function due to finite field effects) affects all measurements in a near identical manner.

\subsection{The Two-Point Correlation Function}

The angular two-point correlation function $\omega(\theta)$ is a measure of how much more likely it is to find two galaxies at a given angular separation than in a uniform  Poissonian distribution:

\begin{equation}
dP=\sigma(1+\omega(\theta))d\Omega ,
\end{equation}

where $dP$ is the probability of finding two galaxies at an angular separation $\theta$, $\sigma$ is the surface number density of galaxies and $d\Omega$ is solid angle. We require $\omega(\theta)>-1$ and $\lim_{\theta \to \infty}\omega(\theta) = 0$ for non-negative probabilities and for non-infinite surface densities respectively. The two-point correlation function is constructed to not be a function of number density (although you might expect them to be related for physical reasons).

The conventional way to estimate $\omega(\theta)$ for the auto-correlation function is with the \citet{Landy1993} estimator, which is based on calculating $DD(\theta)$, the normalised number of galaxies at a given separation in the real data, $RR(\theta)$, the corresponding figure for a synthetic catalogue of random galaxies identical to the data catalogue in every way (i.e. occupying the same field) except position, and $DR(\theta)$, the number of galaxy to synthetic point pairs:

\begin{equation}
\omega(\theta)=\frac{DD-2DR+RR}{RR} .
\end{equation}

\subsection{Halo Occupation Distribution Modelling}

Halo Occupation Modelling is a popular way of modelling galaxy clustering measurements, and has been shown to give physical results in agreement with other methods (e.g. \citealp{Coupon2015,Chiu2016}, and see \citealp{Coupon2012} and \citealp{McCracken2015} for a more complete breakdown). A given set of galaxy occupation statistics is given, usually parametrised by 3-5 numbers, e.g. the number of galaxies in a halo as a function of halo mass. The model correlation function is broken down to a `1-halo' term, describing the small-scale clustering of galaxies within an individual halo, and a `2-halo' term, describing the clustering of the halos themselves. The `1-halo' term is constructed by convolving the profile of galaxies within a halo with itself, weighting by the number of galaxies in the halo, and then integrating over all halo masses. The profile is usually taken to be one galaxy at the centre of the halo (the `central') and all other galaxies tracing a Navarro-Frank-White (NFW; \citealp{Navarro1996}) profile. The 2-halo term is constructed by scaling the dark matter linear correlation function by the weighted-average halo bias of the host halos. The transition from the non-linear 1-halo term, to the linear 2-halo term, occurs at $\sim 0.02-0.05^{\circ}$ (i.e. within the angular scales we measure clustering for) over the redshifts considered in this work.

The most general HOD parametrisation commonly used is that of \citet{Zheng2005}, that gives the total number of galaxies in a halo as:

\begin{equation}  \label{eq:params_tot}
\langle N_{\rm tot}(M_{h}) \rangle=\langle N_{\rm cen}(M_{h}) \rangle+\langle N_{\rm sat}(M_{h}) \rangle ,
\end{equation}

the total number of central galaxies as:

\begin{equation}  \label{eq:params_cen}
\langle N_{\rm cen}(M_{h}) \rangle=\frac{1}{2} \left(1+ \mathrm{erf} \left( \frac{\log_{10}M_{h}-\log_{10}M_{\rm min}}{\sigma_{\log_{10} M}} \right)  \right) ,
\end{equation}

and the total number of satellites as:

\begin{equation}  \label{eq:params_sat}
\langle N_{\rm sat}(M_{h}) \rangle=\langle N_{\rm cen}(M_{h}) \rangle \left( \frac{M_{h}-M_{0}}{M_{1}} \right)^{\alpha} 
\end{equation}

(when $M_{h}>M_{0}$; otherwise the number of satellites is zero). This model has five parameters; $M_{\rm min}$ describes the minimum halo mass required to host a central galaxy, $\sigma_{\log M}$ describes how sharp this step jump is (equivalent to the central to halo mass scatter), $M_{0}$ is a halo mass below which no satellites are found, and $M_{1}$ is the scale mass at which the halo begins to accumulate satellites. The power law index $\alpha$ describes how the number of satellites grows with halo mass.

\subsection{MCMC Fitting} \label{sec:MCMC_description}

To model our angular correlation functions, we compare to model correlation functions from the {\sc Halomod}\footnote{https://github.com/steven-murray/halomod} code (Murray, Power, Robotham, in prep.). First a spatial correlation function is calculated, which is then projected to an angular correlation function (as per \citealt{Limber:1954zz}), using a redshift distribution derived by smoothing the true redshift values of the galaxies for the $\cal{C}_{\rm Hzagn,sim}$ correlation function, and the weighting system of \citet{Arnouts2002} for the $\cal{C}_{\rm Hzagn,obs}$ correlation function. We then subtract off the numerical approximation of the integral constraint to get our final model correlation function.

We use {\sc Emcee}\footnote{http://dan.iel.fm/emcee/current/} (\citealp{Foreman-Mackey2012})  to provide a Markov Chain Monte Carlo sampling of the parameter space to fit our correlation function. The  likelihood is defined by:

\begin{equation}  \label{eq:chi}
\chi^{2}= \frac{[n_{\mathrm{\mathrm{gal}}}^{\mathrm{obs}}-n_{\mathrm{\mathrm{gal}}}^{\mathrm{model}}]^{2}}{\sigma_{n}^{2}}  +  \sum\limits_{i} \frac{[\omega^{\mathrm{obs}}(\theta_{i})-\omega^{\mathrm{model}}(\theta_{i})]^{2}}{\sigma_{w_{i}}^{2}} ,
\end{equation}

where $n_{\mathrm{\mathrm{gal}}}^{\mathrm{obs}}$ is the $\cal{C}_{\rm Hzagn,sim}$/$\cal{C}_{\rm Hzagn,obs}$ galaxy number density, $n_{\mathrm{\mathrm{gal}}}^{\mathrm{model}}$ is the HOD model galaxy number density, $\sigma_{ n}$ is the error on the  number density including both Poisson noise and cosmic variance (calculated as per \citealp{Trenti2008}), $\theta_{i}$ are the angular scales we fit over, $\omega^{\mathrm{obs}}$ is the observed angular correlation function, $\omega^{\mathrm{model}}$ is the angular correlation function of a given HOD model, and $\sigma_{\omega_{i}}$ is the error on the measurements of the correlation function from bootstrapping of the data.

We use a uniform prior over $10<\log_{10}{(M_{\mathrm{min}}/\mathrm{M}_{\odot})}<15$, $\log_{10}{(M_{\mathrm{min}}/\mathrm{M}_{\odot})}<\log_{10}{(M_{\mathrm{1}}/\mathrm{M}_{\odot})}<17$, $8<\log_{10}{(M_{0}/\mathrm{M}_{\odot})}<\log_{10}{(\mathrm{M}_{\mathrm{1}}/\mathrm{M}_{\odot})}$ (uniform in log space), $0.5<\alpha<2.5$ and $0<\sigma<0.6$.  We used 20 walkers with 1000 steps, which have starting positions drawn uniformly from the prior.

We use 500,000 random data points in this study. We use 100 bootstrap resamplings to estimate the uncertainty at the 16th and 84th percentiles of the resampling. For $n_{\mathrm{\mathrm{gal}}}^{\mathrm{obs}}$, for $\cal{C}_{\rm Hzagn,sim}$ we use the number of galaxies divided by the appropriate volume, for $\cal{C}_{\rm Hzagn,obs}$ we use the sum of the weights as for $\cal{C}_{\rm VIDEO}$ \citep{Hatfield2016}.

\section{Results} \label{sec:horizon_results}

\subsection{ACF Measurements in Horizon-AGN}  \label{sec:horizon_acfs}
\begin{figure*}
\centering
\includegraphics[scale=0.6]{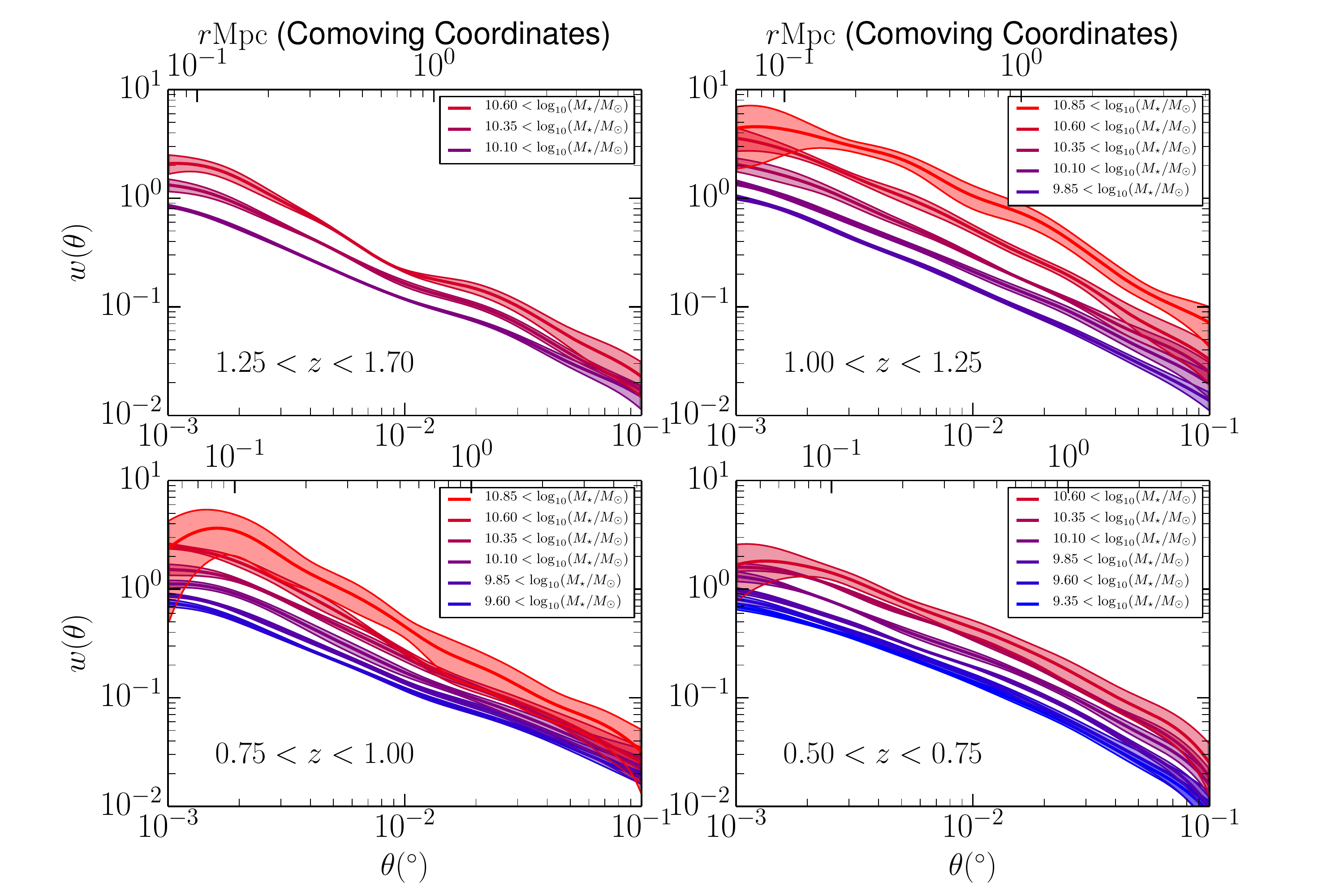}
\caption{The angular correlation function for each stellar mass and redshift bin, measured in the simulated catalogue $\cal{C}_{\rm Hzagn,sim}$ in different redshift bins. The dotted lines represent 1-$\sigma$ uncertainties on the measurements. The comoving scale on the upper x-axes correspond to the median redshift of the samples.}
\label{fig:horizon_mock_acfs}
\end{figure*}

The clustering measurements from $\cal{C}_{\rm Hzagn,sim}$, i.e without any observational errors, are shown in Fig.~\ref{fig:horizon_mock_acfs}. 
The behaviour usually measured in observational datasets is correctly recovered. 

The mock correlation functions have near power-law behaviour, with sensible amplitudes increasing with stellar mass at all redshifts. To some degree this is expected - the underlying dark matter distribution is robust between different simulations, and bias is expected to be $\lesssim 2-3$, so it would be extremely surprising if it was orders of magnitude greater or less than expected. But the agreement over linear and non-linear scales should reassure us that {\sc Horizon-AGN} is correctly qualitatively capturing the large-scale distribution of galaxies. The clustering measurements from $\cal{C}_{\rm Hzagn,obs}$ look very similar; how they compare to $\cal{C}_{\rm Hzagn,sim}$ is discussed in section \ref{sec:observed_versus_true_acfs}. The exception is the highest stellar mass bin at $0.75<z<1$ in the LePhare mock catalogue, for which there were too few galaxies to measure the clustering - this $\cal{C}_{\rm Hzagn,obs}$ bin is excluded in the subsequent analyses.

\subsection{Impact of Observational Uncertainties on Clustering Measurement}\label{sec:observed_versus_true_acfs}

\begin{figure*}
\centering
\includegraphics[scale=0.6]{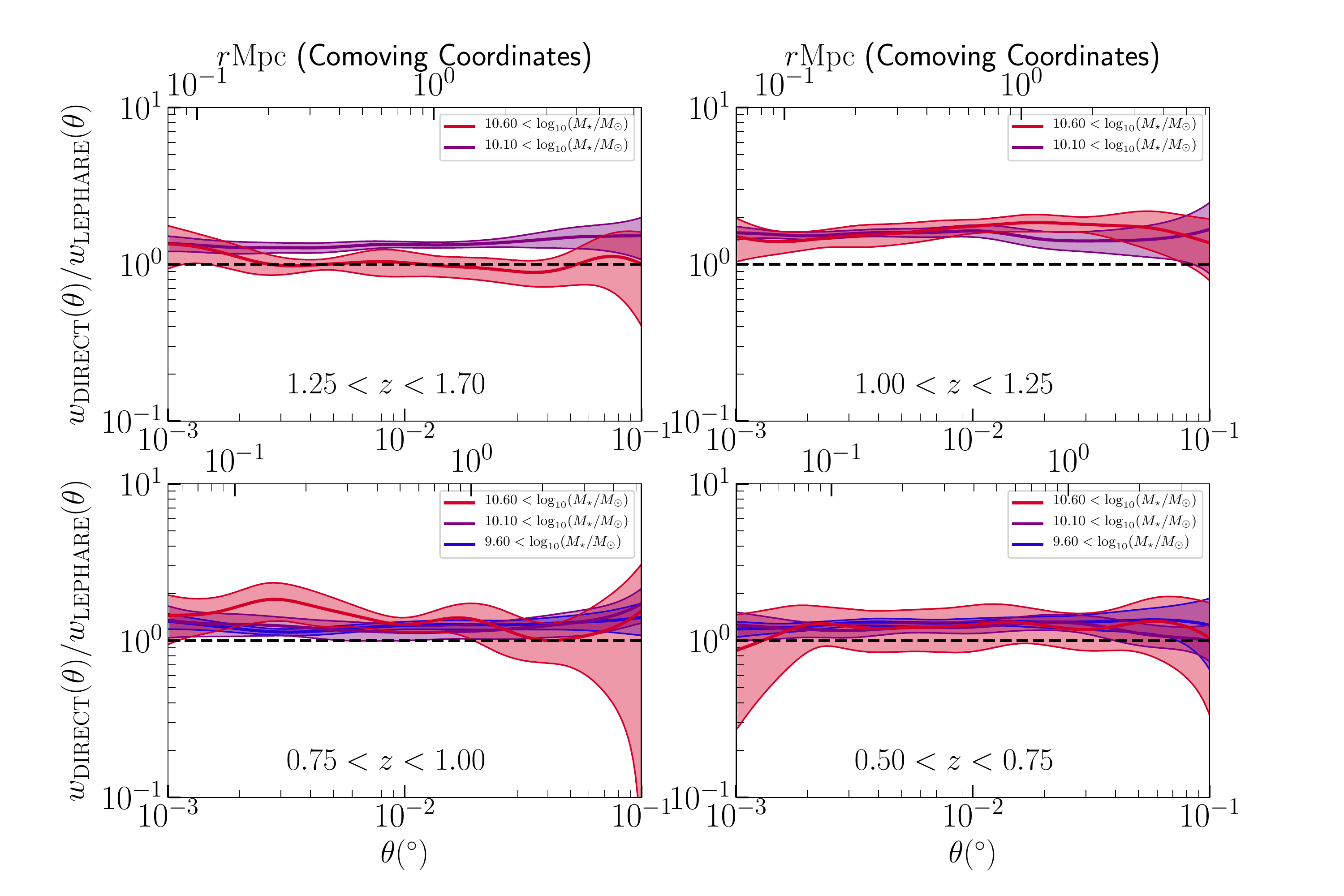}
\caption{Ratio of the {\sc Horizon-AGN} clustering measurements for the stellar mass and redshift bins when a) taking the stellar masses and redshifts directly from the simulation ($\cal{C}_{\rm Hzagn,sim}$) and b) taking the stellar masses and redshifts from the LePhare based catalogues ($\cal{C}_{\rm Hzagn,obs}$). Only alternating stellar mass bins plotted for clarity of plot. The thin lines are the errors on the ratio, with the errors from both sets of {\sc Horizon-AGN} clustering measurements propagated through.The dashed line corresponds to $\omega_{\mathrm{DIRECT}}=\omega_{\mathrm{LEPHARE}}$. }
\label{fig:horizon_direct_lephare_compare}
\end{figure*}

In Fig. \ref{fig:horizon_direct_lephare_compare} we compare the clustering measurements from the mock catalogues $\cal{C}_{\rm Hzagn,sim}$ and $\cal{C}_{\rm Hzagn,obs}$ for the same stellar mass and redshift bins when the sources are binned  by a) their true redshift and stellar mass within the simulation and b) their photometric redshift and stellar mass (as derived from the simulated fluxes, and using the same weighting scheme as was applied to $\cal{C}_{\rm VIDEO}$) respectively.

The two sets of clustering measurements agree to moderate accuracy, never differing by more than a factor of two, typically with comparatively little angular dependence. The $\cal{C}_{\rm Hzagn,sim}$ clustering measurement is also consistently higher than inferred from $\cal{C}_{\rm Hzagn,obs}$ because of scattering between stellar mass and redshift bins, which inevitably reduces the clustering signal. Although weak, there is slight trend for this systematic reduction to increase with increasing redshift, as redshift uncertainties tend to be higher at higher redshift.

\subsection{Comparing VIDEO and {\sc Horizon-AGN}}

\begin{figure*}
\centering
\includegraphics[scale=0.6]{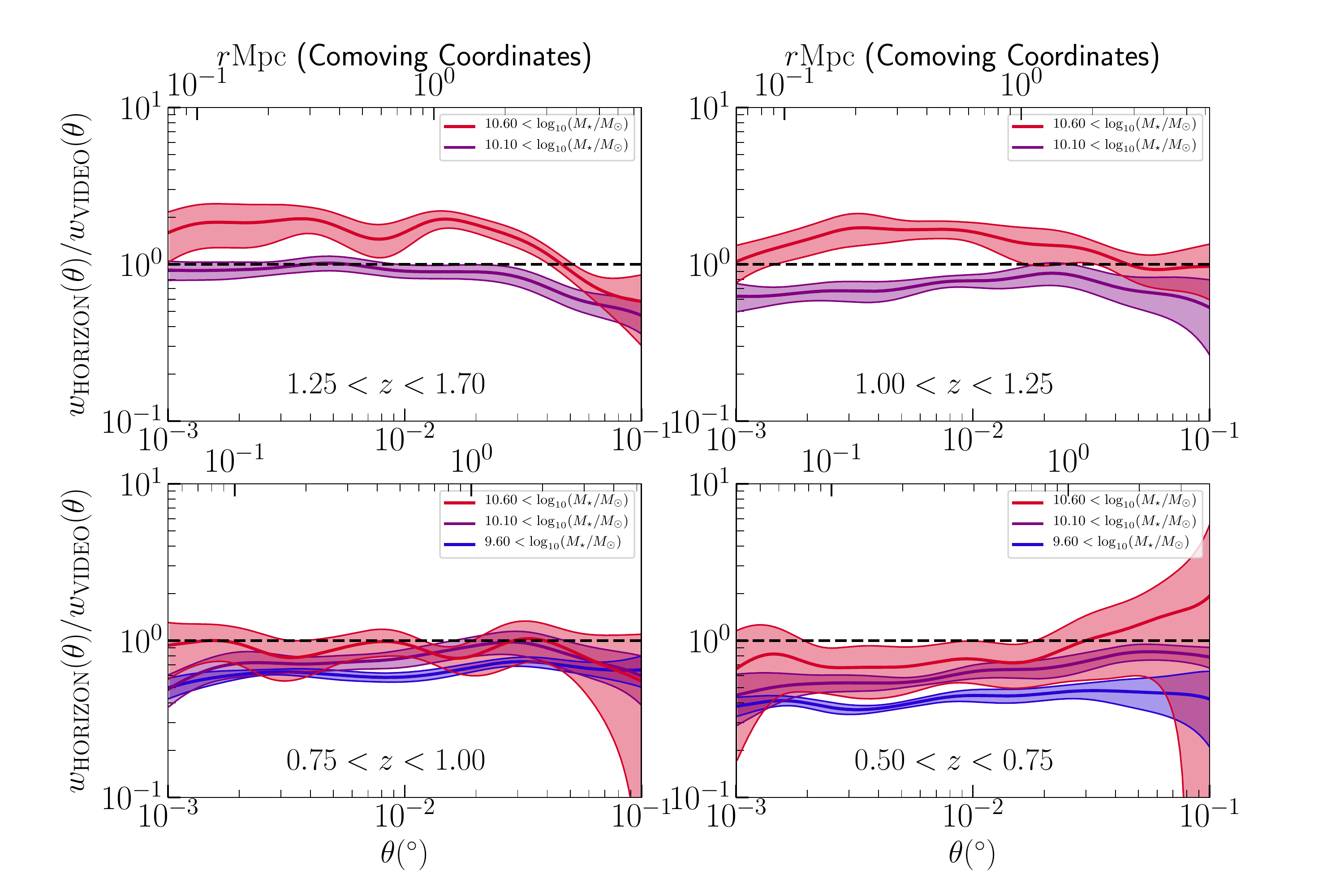}
\caption{Ratio of the VIDEO and $\cal{C}_{\rm Hzagn,obs}$ clustering measurements for the stellar mass and redshift bins. Only alternating stellar mass bins are plotted for clarity of plot. The shaded area covers the uncertainties on the ratio, with the uncertainties on both the VIDEO and {\sc Horizon-AGN} measurements propagated through. The dashed line corresponds to $\omega_{\mathrm{HORIZON}}=\omega_{\mathrm{VIDEO}}$. }

\label{fig:horizon_VIDEO_compare}
\end{figure*}

We compare the $\cal{C}_{\rm Hzagn,obs}$ results to the corresponding VIDEO sub-sample measurements from \citet{Hatfield2016}. In Fig. \ref{fig:horizon_VIDEO_compare} we compare the $\cal{C}_{\rm Hzagn,obs}$ clustering measurements, as opposed to the $\cal{C}_{\rm Hzagn,sim}$ clustering measurements in order to be comparing `like-for-like' since the VIDEO results are also derived from photometry and have similar biases. In particular we show the ratio of the VIDEO and {\sc Horizon-AGN} angular correlation functions in Fig.~\ref{fig:horizon_VIDEO_compare}. Note that the 1-point statistics of these samples, at the same stellar masses, are quite different (figure \ref{fig:number_counts}), which should be borne in mind when interpreting this data.

The {\sc Horizon-AGN} clustering measurements are greater or lesser than the VIDEO measurements by at most a factor of 3 in any bin. In general the trend in each redshift bin is that  $\omega_{\mathrm{HORIZON}}/\omega_{\mathrm{VIDEO}}$ increases with stellar mass e.g. at lower masses {\sc Horizon-AGN} increasingly underestimates the clustering, indicating that clustering in {\sc Horizon-AGN} has too strong a dependence on stellar mass. At lower redshifts {\sc Horizon-AGN} underestimates the clustering, and at higher redshifts is closer to the VIDEO clustering, sometimes starting to slightly overestimate it. The underestimation (particularly at low redshifts and stellar masses) has very high statistical significance ($>5\sigma$) , however the over-estimations (at higher stellar masses and redshifts) typically have lower (1-2$\sigma$) significance. In most bins there is very little angular scale dependency with the possible minor exception of large scales at high redshift - possibly a consequence of that bin in {\sc Horizon-AGN} having low power at large scales.

\subsection{HOD Fits} \label{sec:horizon_fits}

\begin{figure*}
\centering
\includegraphics[scale=0.8]{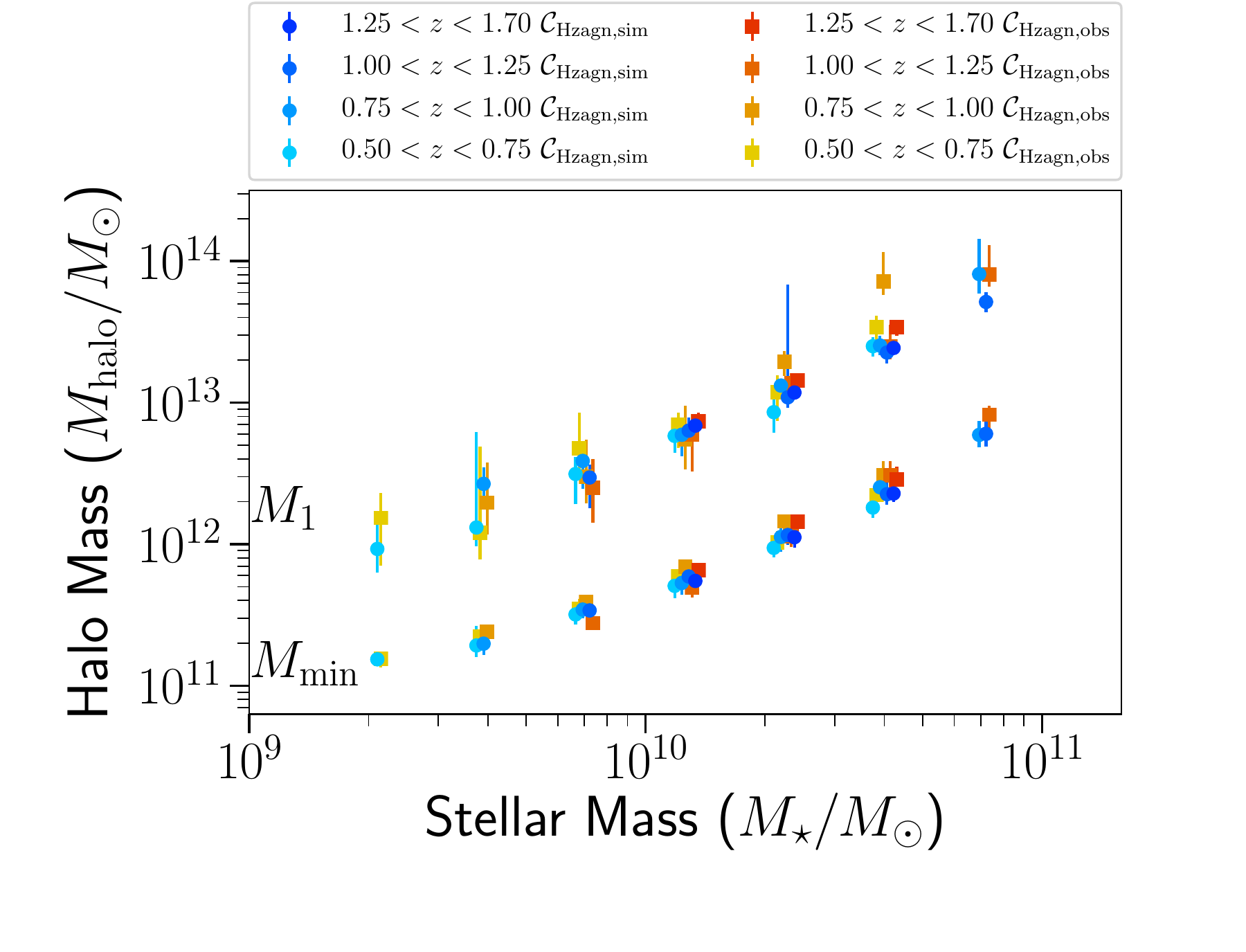}
\caption{Best fit $M_{\rm min}$ and $M_{1}$ for the {\sc Horizon-AGN} clustering measurements, as a function of galaxy sample threshold stellar mass, for both $\cal{C}_{\rm Hzagn,sim}$  and $\cal{C}_{\rm Hzagn,obs}$. Data is plotted with small stellar mass offsets for clarity of plot.}
\label{fig:horizon_direct_lephare_hod_params}
\end{figure*}

Fig.~\ref{fig:horizon_direct_lephare_hod_params} shows the HOD $M_{\rm min}$ and $M_{1}$ parameters (characteristic halo masses required to host central and satellite galaxies respectively). This is plotted as a function of stellar mass for the different redshift bins, as inferred from $\cal{C}_{\rm Hzagn,sim}$ to from $\cal{C}_{\rm Hzagn,obs}$ (both in {\sc Horizon-AGN}). The parameters are largely very similar when inferred from the direct catalogue and the LePhare catalogue. We do not show the plots here, but the $\alpha$ values for both catalogues agree within uncertainties, and  are essentially consistent with $\alpha=1$ (as per the results in \citealp{Hatfield2016} for the VIDEO observations). The $M_0$ values also agree, but again similarly to \citet{Hatfield2016} are very poorly constrained (as it describes a cut-off in the number of satellites at low halo masses where the expected number of satellites is already much less than one).

\begin{figure*}
\centering
\includegraphics[scale=0.8]{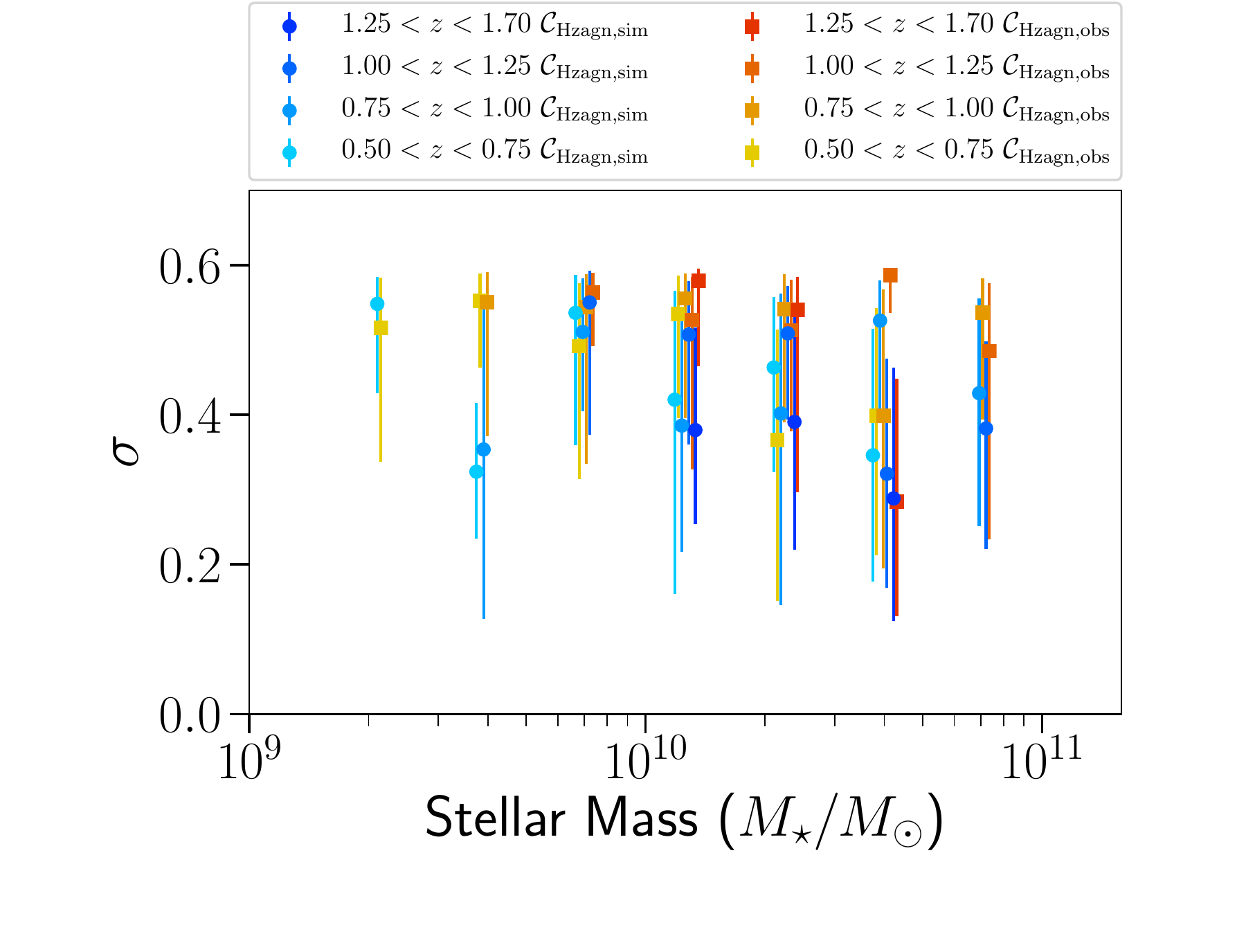}
\caption{Best fit $\sigma$ values for the {\sc Horizon-AGN} clustering measurements, as a function of galaxy sample threshold stellar mass, for both $\cal{C}_{\rm Hzagn,sim}$and $\cal{C}_{\rm Hzagn,obs}$. Data is plotted with small stellar mass offsets for clarity of plot.}
\label{fig:horizon_direct_lephare_hod_params_sigma}
\end{figure*}

Fig.~\ref{fig:horizon_direct_lephare_hod_params_sigma} shows the HOD $\sigma$ parameter as a function of stellar mass for the different redshift bins, as inferred from $\cal{C}_{\rm Hzagn,sim}$ compared to those from $\cal{C}_{\rm Hzagn,obs}$ (both in {\sc Horizon-AGN}). The scatter inferred when observational uncertainties are included is significantly higher: $\sim0.6$ typically measured for $\cal{C}_{\rm Hzagn,obs}$ and $\sim0.3-0.5$ typically measured from $\cal{C}_{\rm Hzagn,sim}$. Note the use of a uniform prior over [0,0.6] for $\sigma$ as per \citet{McCracken2015} and \citet{Hatfield2016} is now substantiated by the knowledge that observational systematics bias scatter estimates to higher values. The true scatter in the simulation is even lower, $\sim0.2$, so even without the observational systematics we still overestimate the scatter by about 50\% through the fitting process.

\begin{figure*}
\centering
\includegraphics[scale=0.8]{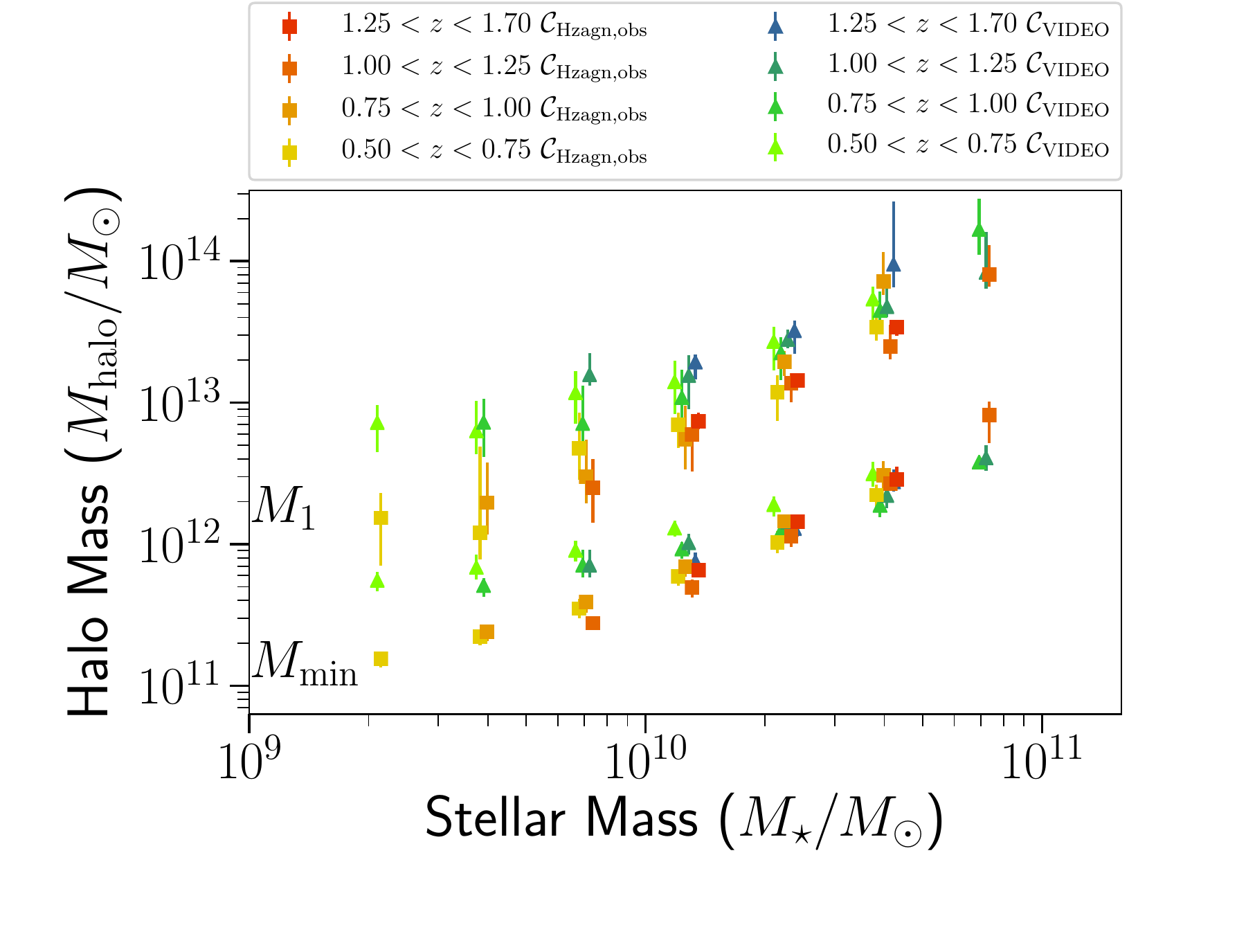}
\caption{Best fit $M_{\rm min}$ and $M_{1}$ for the {\sc Horizon-AGN} $\cal{C}_{\rm Hzagn,obs}$ clustering measurements, as a function of galaxy sample threshold stellar mass. Equivalent measurements in VIDEO from \citet{Hatfield2016} shown for comparison. Data is plotted with small stellar mass offsets for clarity of plot.}
\label{fig:horizon_hod_params}
\end{figure*}

We show in Fig.~\ref{fig:horizon_hod_params} the best fit $M_{\rm min}$ and $M_{1}$ values from the HOD models fit to the LePhare {\sc Horizon-AGN} measurements, alongside the equivalent VIDEO measurements from \citet{Hatfield2016}. The {\sc Horizon-AGN} fits give the same qualitative behaviour as observed in VIDEO; $M_{\rm min}$ and $M_{1}$ growing as approximate power laws with stellar mass, with $M_{1}$ roughly an order of magnitude larger than $M_{\rm min}$. {\sc Horizon-AGN} measurements, for both $M_{\rm min}$ and $M_{1}$ are generally slightly lower than the equivalent VIDEO measurements.  $M_{\rm min}$ values agree well at high stellar masses ($\sim 10^{10.85} M_{\sun}$), but {\sc Horizon-AGN} $M_{\rm min}$ values are around 0.5 dex lower than the VIDEO values for stellar masses $\sim 10^{9.6} M_{\sun}$. This is consistent with Fig.~\ref{fig:horizon_VIDEO_compare}, where the clustering amplitude in {\sc Horizon-AGN} is much lower than that measured in VIDEO at lower stellar masses. $M_{1}$ values are consistently a factor of $\sim 2$ lower for {\sc Horizon-AGN} than for VIDEO. Similarly to in our VIDEO measurements, there appears to be very little evolution in $M_{\rm min}$ and $M_{1}$ as a function of redshift, over the redshift and stellar mass ranges considered here.

\section{Discussion} \label{sec:horizon_discussion}

Causes of discrepancies between observed and simulated measurements can be divided into three categories. Firstly, discrepancies can arise from the simulation inherently not completely capturing the correct galaxy formation physics e.g. the simulation genuinely creates more galaxies of a given mass than are present in the real Universe. Secondly, discrepancies can arise when observational biases are poorly understood and accounted for e.g. the incompleteness of a galaxy survey is underestimated, leading us to falsely conclude that the simulation over produces galaxies, even if the simulation actually matches reality. Finally, there can be differences greater than expected from statistical uncertainty from \textit{cosmic variance} - the galaxy formation physics could be correctly captured in the simulation, but large-scale structure variance could still make measurements from the simulation differ from observations non-trivially.

\subsection{Observational Biases} \label{sec:observational_biases}

A key complexity in comparing results from observations and simulations is that the two `tools' perceive galaxies in a dramatically different manner. Simulations are concerned with (stellar) mass, whereas observations deal with luminosity. Comparisons between observations and simulations inherently require either inferring mass from observed luminosities, or predicting a luminosity from a simulated mass. 

\cite{Laigle2019} fit a library of templates to the {\sc Horizon-AGN} mock photometry (essentially treating the photometry as if it was observations) to investigate if the true galaxy masses and redshifts within the simulation can be recovered correctly from the photometry. In the COSMOS configuration, they find a scatter of $\sim0.1$~dex between the intrinsic and photometric stellar masses (see their Fig.~6). More problematically, they also find (for the redshifts relevant to this work) that the inferred stellar masses are biased to values at most $\sim0.12$~dex less than the true values. This estimate holds for the COSMOS configuration, i.e. for redshifts and masses derived from the fitting of 30 photometric bands. However, even if the photometric baseline used in our analysis contains less photometric bands, the NIR photometric range (which matters the most for stellar mass computation) is still very well sampled with the VIDEO observations. Therefore the performance of the VIDEO configuration in terms of the precision of the reconstructed stellar mass is very similar to the COSMOS one.  Clearly the systematic underestimation of the mass is relevant for consistently comparing clustering results between observations and simulations. If it is the case that observationally inferred stellar masses are consistently 0.1 dex less than the true values, then comparing the clustering of $M_{\star}>10^{10} M_{\sun}$ observed galaxies with clustering of $M_{\star}>10^{10} M_{\sun}$ simulated galaxies is really comparing the clustering of $M_{\star}>10^{10.1} M_{\sun}$ observed galaxies to $M_{\star}>10^{10} M_{\sun}$ simulated galaxies. The higher mass sample will typically have a higher clustering amplitude, so underestimating the stellar masses of observed galaxies will falsely make it appear that the simulation underestimates clustering amplitude (even if it correctly captures galaxy physics). Scatter in the stellar mass conversely will reduce clustering amplitude, as the more numerous lower stellar mass galaxies (in lower mass haloes) will be preferentially `up-scattered' into higher stellar mass bins. Photometric redshift uncertainties will ordinarily have comparatively little impact when the scatter is within an individual redshift bin, but scatter between redshift bins will typically reduce the clustering amplitude, as structure at redshifts with large separations is largely uncorrelated.

Comparison of the HOD fits from the $\cal{C}_{\rm Hzagn,sim}$ and $\cal{C}_{\rm Hzagn,obs}$ catalogues allows us to investigate directly the impact of using photometric stellar masses and redshifts instead of the true value, and develops the discrepancy highlighted in Fig.~\ref{fig:horizon_direct_lephare_compare}. In particular the scatter in stellar mass itself as the  induces a higher measured $\sigma$ (effectively  scatter in central stellar mass to halo mass ratio) -- the $\cal{C}_{\rm Hzagn,obs}$ scatter is about 0.1-0.2 dex greater than the $\cal{C}_{\rm Hzagn,sim}$ scatter (see Fig.~\ref{fig:horizon_direct_lephare_hod_params_sigma}), which is comparable to the scatter in Fig.~6 in \citet{Laigle2019}. In other words the measured scatter in stellar mass to halo mass ratio should be interpreted as the sum of both the intrinsic and the photometric scatter. The $\sigma$ measured in VIDEO observations in \citet{Hatfield2016} is also $\sim0.6$ - our results would suggest that this is likely an overestimate resulting from scatter in the estimates of the stellar mass. 

Similarly the marginally higher $M_{\rm min}$ values for $\cal{C}_{\rm Hzagn,obs}$ corresponds to the bias in the estimate of the stellar mass; the $\cal{C}_{\rm Hzagn,obs}$ points should essentially be slightly shifted to the right in Fig. \ref{fig:horizon_direct_lephare_hod_params} and \ref{fig:horizon_hod_params}. In summary, the reduced clustering amplitude from photometric systematics described in section~\ref{sec:observed_versus_true_acfs} mainly increases our measurements of the SMRH scatter in HOD modelling, as opposed as biasing the estimated halo masses themselves. 

\paragraph*{Limitations of our modelling}

Our estimates of the systematics in the amplitude of the clustering must be taken as an optimistic lower limit. As briefly described in Section~\ref{sec:simulation_description} and in more details in \cite{Laigle2019},  there are two main limitations of our modelling of observational uncertainties.
 
Firstly, the mock photometry of the simulated galaxies includes less variety than the real photometry (mainly due to the use of constant IMF and SPS models, a constant dust-to-metal ratio and single dust attenuation law, and no modelling of the emission lines). For example, fitting the photometry with a different IMF as the one used to produce this photometry might cause additional systematics of the order of 0.1 dex\footnote{Magnitudes are $\sim 0.4$~mag fainter with a Salpeter IMF compared to a Chabrier one, so a mis-match between both IMF at the SED-fitting stage would cause a 0.1 offset in stellar mass.}.  
 
 Secondly, although photometric errors are included in a statistical sense in our mock photometry, we do not model any of the possible systematics occurring at the stage of photometric extraction. These systematics are numerous and might include, in particular: PSF variations over the field and as a function of wavelengths, blending of nearby galaxies or conversely fragmentation of galaxies with perturbed morphologies, errors in the astrometry, imperfect removal of the background. The only way to quantify consistently the importance of all these additional sources of noise is to perform a end-to-end extraction of the photometry directly from mock images, tuned to incorporate the characteristics of the instrument, which will be presented in a future work. Most of these effects are however likely to be important especially at faint magnitudes, and our cut at $K_{\rm s}<23.5$ should preserve the vast majority of galaxies in our sample from strong systematics. But it must be noted that bright objects might still suffer from some of these effects e.g. PSF variations or miscentering. 
 
 As a consequence of these two limitations in our pipeline, the redshift and masses uncertainties in $\cal{C}_{\rm Hzagn,obs}$ are likely to be underestimated, as outlined by  Fig.~4 in \citealp{Laigle2019} (a comparison of COSMOS2015 and {\sc Horizon-AGN} errors as derived from {\sc LePhare}). We would anticipate that increasing the redshift uncertainties would further increase the underestimation of the clustering amplitude, and therefore the analysis presented here should be considered as an optimistic case.

\subsection{Over Production of Lower Mass Galaxies in {\sc Horizon-AGN}} \label{sec:over_production}

Although observational biases may play some role\footnote{The {\sc Horizon-AGN} Lephare catalogue and VIDEO have similar observational biases, but only to a certain degree, as discussed in Sec.~\ref{sec:observational_biases}.}, the main discrepancy in the galaxy-halo connection from clustering between observations and simulations apparent in our results is that the stellar mass to halo mass ratio (SMHR) doesn't drop off as fast at low masses in {\sc Horizon-AGN} as in VIDEO observations (fig. \ref{fig:horizon_hod_params}).

This raised SMHR for low mass haloes in {\sc Horizon-AGN} (equivalent to the over production of lower mass galaxies in, or galaxies in lower mass haloes growing too large relative to observations) is reported in \citet{Dubois2014} and \citet{Kaviraj2017}. \citet{Kaviraj2017} compare stellar mass functions from the literature to those directly from the simulation, finding an over abundance of galaxies with stellar masses $\lesssim 10^{11} M_{\sun}$ (c.f. figure \ref{fig:number_counts}). \citet{Dubois2014} compare the SMHR for central galaxies and their haloes directly from the simulation, to observational results from abundance matching in \citet{Moster2013}, finding that SMHR observations and simulations agree for haloes with masses $\approx 10^{12} M_{\sun}$, but that galaxies in $\approx 10^{11} M_{\sun}$ haloes are approximately a factor of ten more massive in {\sc Horizon-AGN} than observations would suggest, which agrees with our results. \citet{Kaviraj2017} suggest that this discrepancy is due to missing physics in the sub-grid supernovae feedback prescription of the simulation. They propose that stronger regulation of star-formation in lower mass haloes could be achieved either through more realistic treatment of the interstellar medium (\citealp{Kimm2015}), or through stronger winds from star clusters (\citealp{Agertz2014}). Note that the low halo mass regime is one in which using clustering as opposed to abundance matching is particularly important for inferring galaxy to halo mass ratios (see \citealp{Sawala2014}).

\begin{figure*}
\centering
\includegraphics[scale=0.4]{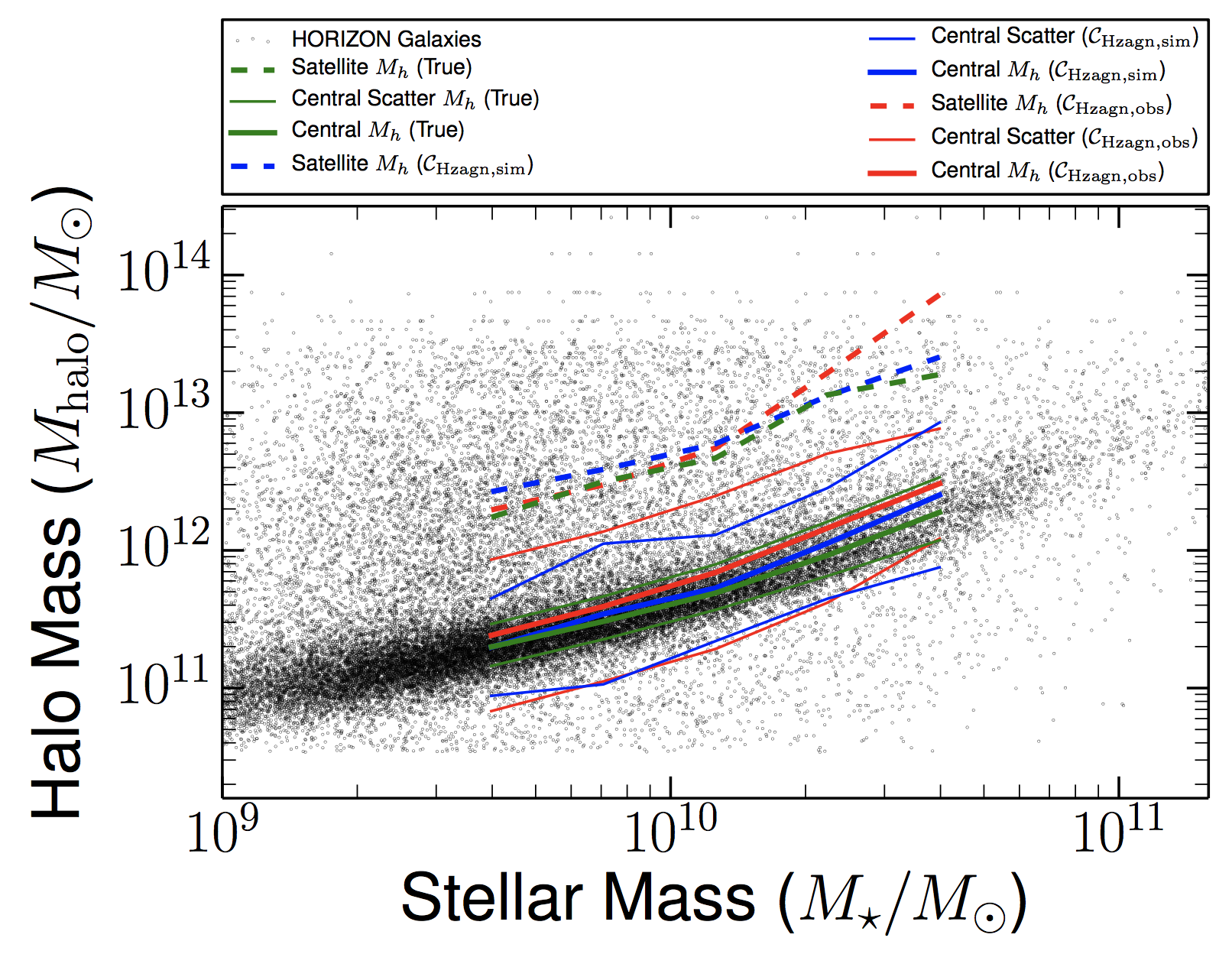}
\caption{Comparison of the stellar mass to host halo mass for galaxies taken directly from {\sc Horizon-AGN} (black points), and stellar mass when central (filled line) with scatter (thin lines) and when satellites (dashed line), as inferred from clustering. The red curves show the inference from $\cal{C}_{\rm Hzagn,obs}$, the blue the curves from $\cal{C}_{\rm Hzagn,sim}$, and the green curves the true values within the simulation. N.B. the scatter shown here is the inferred scatter on the central stellar mass to halo mass relation, not the uncertainty on this relation. }

\label{fig:true_halo_masses}
\end{figure*}

With simulations it is possible to compare inferences made from mock observations to the true physics within the simulation e.g. it is possible to compare the inferred halo masses from clustering to the true halo masses within the simulation. Fig.~\ref{fig:true_halo_masses} shows the HOD inferences compared with the real halo masses in the simulation, and illustrates that the inferences from the clustering and HOD modelling essentially agree with the true galaxy-halo relation in the simulation. This suggests that the modelling in \citet{Hatfield2016} (and other similar works) is largely giving accurate inferences. The bias introduced to the estimates of halo masses from observational systematics is $\sim0.1$ dex (the difference between the inferences based on $\cal{C}_{\rm Hzagn,sim}$ and $\cal{C}_{\rm Hzagn,obs}$), and the bias introduced by the fitting process is $\sim0.05-0.1$dex\footnote{With the exception of $M_1$ for the highest stellar masses, where it gets up to $\sim0.5$dex} (the difference between the inferences from $\cal{C}_{\rm Hzagn,sim}$ correlation functions, and the true physics within the simulation). The physics of the {\sc Horizon-AGN} universe are, as discussed, different to that in the real Universe, so the sizes of systematics are not guaranteed to be exactly the same size for observational data, but this seems a plausible estimate of the sizes of systematics on real observations. This is particularly comforting given the comparative simplicity of HOD models, for example a) the model discussed here doesn't include assembly bias (see \citealp{Zentner2014}) and b) there is some evidence at low redshift that galaxies aren't spherically symmetrically distributed in haloes (\citealp{Pawlowski2017}). \citet{Pujol2017} in contrast found SAMs and HOD models give different clustering predictions, mainly because of contributions of orphan galaxies and because the SAMs had different satellite galaxy radial profile densities than the NFW that are assumed for the HOD model (modifying the profiles is considered in \citealp{Hatfield2017} however). Hydrodynamical simulations do not have orphan galaxies per say, but they do capture some of the physics in a more fundamental manner; HOD models certainly will not capture everything within the simulation, but our results seem to show that they correctly capture the main features.

\subsection{Cosmic Variance}

Cosmic variance refers to the observation that statistical measurements can vary by more than Poisson variance due to large-scale structure e.g. the field might sample an extreme over- or under-density of the Universe which can impact on the statistical properties of the field. As discussed in \citet{Hatfield2016}, there is already known to be moderate difference in the clustering properties of the VIDEO/CFHTLS-D1 field and the COSMOS field (\citealp{Durkalec2015}).

Cosmic variance-like effects, as described in \citet{Blaizot2005}, can also impact mock observations derived from simulations in two main ways. Firstly only a finite volume of the Universe is simulated. This requires sampling initial conditions from some larger distribution - a different sampled set of initial conditions would lead to a different simulated universe. Secondly through the construction of the mock cone; for a given simulation, constructing the mock sky `from different viewpoints' will give different sets of mock observations.

VIDEO has three separate fields, so in future work it should be possible to directly estimate the effects of cosmic variance on our observational results. To understand cosmic variance in the simulated results, ideally one would run the simulation multiple times, but this is currently unfeasible for hydrodynamical simulations\footnote{Although \citet{Pontzen2016} present an interesting way of capturing much of cosmic variance in hydrodynamical simulations using only \textit{2} runs.} (although it is more viable in SAMs e.g. \citealp{Stringer2008a}).  The approach of sampling different viewpoints within the same simulation (the second type of cosmic variance in simulations discussed above) may be more viable for hydrodynamic simulations, although as \citet{Blaizot2005} discuss, this will generally underestimate the true cosmic variance, as each generated mock sky will still be sampling the same density field, just from different angles.  

Although to completely quantify cosmic variance multiple observational fields and multiple simulations with different initial conditions are needed, we can estimate the approximate size of cosmic variance if a few assumptions are made. In particular, as discussed, the galaxy-halo connection appears to have relatively little evolution with redshift. If we assume that the galaxy-halo connection is constant in all our redshift bins, and in addition we find it acceptable to treat the different redshift bins as independent volumes of space, we can use the differences between different redshift bins to estimate the approximate size of the cosmic variance (as an alternative to using different fields). Different redshift bins are not completely independent i) because of cross-bin contamination in the catalogues with observational biases, and ii) because adjacent redshift bins are essentially adjacent comoving volumes, and iii) these comoving volumes are all ultimately drawn from the same simulation. Nonetheless, making these assumptions lets us obtain a lower bound on the size of the cosmic variance on the parameters. We assume that the (non-systematic) uncertainty on our estimates of the HOD parameters can be expressed $\sigma_{\mathrm{Total}}^2=\sigma_{\mathrm{CV}}^2 + \sigma_{\mathrm{Stat}}^2$, where $\sigma_{\mathrm{Total}}$ is the total uncertainty on the estimate, $\sigma_{\mathrm{CV}}$ is the cosmic variance on the estimate, and $\sigma_{\mathrm{Stat}}$ is the statistical uncertainty on the estimates calculated in section \ref{sec:horizon_results}. We then fit (using MCMC) a second order-polynomial in log-log space of the  $\cal{C}_{\rm Hzagn,sim}$\footnote{The final estimate of rough size of the impact of cosmic variance is not strongly affected by using a different catalogue or slightly different ways of fitting the data etc.} $M_{\mathrm{min}} $ estimates as a function of $M_{\star}$. When calculating the liklihood we use Gaussian uncertainties of $\sigma_{\mathrm{Total}}$, where the $\sigma_{\mathrm{Stat}}$ values are as calculated, and $\sigma_{\mathrm{CV}}$ is unknown. This gives four unknowns; the three parameters of the polynomial model, and $\sigma_{\mathrm{CV}}$ - and we find $\sigma_{\mathrm{CV}}\gtrapprox 0.05$dex. This is an imperfect model, as cosmic variance is correlated within and between redshift bins, but likely gives a lower bound of the approximate size of the variance, in the absence of multiple fields and multiple simulations to analyse. We can also compare this estimate of the size of cosmic variance on our HOD measurements, to the cosmic variance on 1-point statistics, which we calculated using the method of \citealp{Trenti2008} for equation \ref{eq:chi}. This approach uses assumptions about the 2-point statistics (e.g. galaxy bias) to make estimates of the size of the cosmic variance of the 1-point statistics. This method estimates the size of the cosmic variance on the counts to be $\sim0.05$dex (slightly variable depending on redshift range under consideration etc.). The cosmic variance on the clustering would depend on the 3-point statistics, so will not be the same as the cosmic variance on the counts, and in addition uncertainty on the clustering does not translate linearly to uncertainty on HOD parameters. Nonetheless the fact that the first method gave a result comparable to the analytic result for number counts, would suggest that $\sim0.05$ dex is probably a reasonable approximation of the size of cosmic variance on halo mass estimates.

\section{Conclusions}

In this paper we investigated the clustering of galaxies in mock catalogues from the hydrodynamic cosmological simulation {\sc Horizon-AGN}, to compare to observations, and to confirm that HOD methods give correct physical deductions. We selected mock sources in the simulation that permitted direct comparison to VIDEO sub-samples described in \citet{Hatfield2016}, using redshifts and stellar masses taken both directly from the simulation, as well as derived from the mock photometry. We then measured the angular correlation function, and fitted HOD models using identical procedures to that applied to the observations, so that a consistent comparison could be made.

We found that the correlation function measurements from the simulations were in qualitative agreement with the correlation functions observed in VIDEO, producing the correct approximate power law behaviour, with the correct stellar mass dependence, and the amplitude differing by at most a factor of 3. HOD modelling of the {\sc Horizon-AGN} measurements also qualitatively recovers the broad nature of the connection between galaxies and haloes suggested by VIDEO, namely $M_{\rm min}$ and $M_{1}$ growing as power laws (with a factor of approximately an order of magnitude between them).

The (central) SMHR differs between {\sc Horizon-AGN} and VIDEO, reinforcing the known result of {\sc Horizon-AGN} that at low halo masses galaxies are more massive than observations would suggest (equivalently the stellar mass function over estimates the number of lower mass galaxies). HOD modelling recovers the galaxy-halo connection as taken directly from the simulation, suggesting that conventional HOD methodology is giving physically correct results.

Our key results are:
\begin{itemize}
  \item The halo masses inferred from HOD modelling of clustering in mock {\sc Horizon-AGN} catalogues closely matches halo masses taken from the simulation directly, justifying our confidence that inferences from HOD modelling in observations are correct
  \item Use of photometric redshifts and stellar masses biases clustering measurements (in particular reducing the amplitude of the angular correlation function by up to a factor of two). This in turn biases the measurement of HOD parameters, with a relatively small impact on the inferred halo masses (essentially shifting the SMHR by the bias on the inferred stellar mass), but causing significant overestimates of the scatter in the inferred central stellar mass to halo mass ratio. Non-statistical uncertainties on halo mass estimates in this study were  $\gtrapprox$0.05 dex from cosmic variance (dependent on field geometry), $\sim$0.1 dex from observational systematics (dependent on photometry used and data reduction process), and $\sim$0.05-0.1 dex from systematics in the fitting process (dependent on modelling used).
  \item Clustering measurements in {\sc Horizon-AGN} and the VIDEO survey observations disagree in a way consistent with the known over-production of low mass ($M_{\star}\lesssim 10^{10.5}\mathrm{M}_{\sun}$) galaxies in {\sc Horizon-AGN}, but agree for other stellar masses.
\end{itemize}

The analysis presented in this paper is based on 1deg$^2$, and as we discussed, cosmic variance could play an important role in driving the discrepancy between the simulation and VIDEO. In order to test how much VIDEO is sensitive to cosmic variance, future work will present an analysis based on the full 12deg$^2$ of the survey. In terms of our understanding of galaxy physics, future work will also investigate the potential need for stronger regulation of star formation in low mass haloes in {\sc Horizon-AGN}, or what other modifications to the simulation could help reduce the discrepancy between clustering observations and predictions. At the high-mass end, as suggested in \citet{Hatfield2017}, AGN feedback likely impacts the small-scale galaxy cross-correlation function. Choice of AGN feedback prescription in Horizon-AGN could play a role in driving the discrepancy with VIDEO data - to test this we would measure the clustering in the twin simulation without AGN feedback.

Finally we will look towards using analyses like that presented in this text to understand how to best make inferences from non-linear galaxy clustering in the future using \textit{Euclid} and LSST. The huge amounts of data collected in these surveys will mean that systematic uncertainties will dominate over statistical uncertainties and understanding observational biases like those explored here will be essential for making the most of these planned facilities in the coming years.

\section*{Acknowledgements}

Many thanks to the anonymous referee whose comments have greatly improved the quality of this paper. The first author wishes to acknowledge support provided through an STFC studentship. Many thanks to Jeremy Blaizot for comments and input to the ideas in this paper and to Steven Murray for advice for using {\sc Halomod}. CL is  supported by a Beecroft Fellowship. JD  acknowledges funding support from Adrian Beecroft, the Oxford Martin School and the STFC. OI acknowledges the funding of the French Agence Nationale de la Recherche for the project `SAGACE'. We thank  Stephane Rouberol  for running the Horizon cluster hosted by the Institut d'Astrophysique de Paris. This work was supported by the Oxford Centre for Astrophysical Surveys which is funded through generous support from the Hintze Family Charitable Foundation, the award of the STFC consolidated grant  (ST/N000919/1), and the John Fell Oxford University Press (OUP) Research Fund. This research was also supported in part by the National Science Foundation under Grant No. NSF PHY-1748958. Based on data products from observations made with ESO Telescopes at the La Silla or Paranal Observatories under ESO programme ID 179.A- 2006. Based on observations obtained with MegaPrime/MegaCam, a joint project of CFHT and CEA/IRFU, at the Canada-France-Hawaii Telescope (CFHT) which is operated by the National Research Council (NRC) of Canada, the Institut National des Science de l'Univers of the Centre National de la Recherche Scientifique (CNRS) of France, and the University of Hawaii. This work is based in part on data products produced at Terapix available at the Canadian Astronomy Data Centre as part of the Canada-France-Hawaii Telescope Legacy Survey, a collaborative project of NRC and CNRS.

\bibliographystyle{mn2e_mod}
\bibliography{HORIZON_paper}

\bsp

\label{lastpage}

\end{document}